# From x∗y=k to Uniswap Hooks: A Comparative Review of Decentralized Exchanges (DEX)

Mohammad Ali Asef [a], Seyed Mojtaba Hosseini Bamakan [b, 1]

[a] Department of Computer Engineering, University of Tehran
[b] Department of Management and Data Science, Yazd University

*Abstract*—**Decentralized Exchanges (DEXs) are pivotal applications in the Decentralized Finance (DeFi) landscape, aiming to facilitate trustless cryptocurrency trading by relying on smart contracts and blockchain networks. The developments in the DEXs sector began with the implementation of an Automated Market Maker (AMM) system using a simple math formula by Uniswap in 2018. Absorbing significant funding and the attention of web3 enthusiasts, DEXs have seen numerous advancements in their evolution. A notable recent advancement is the introduction of hooks in Uniswap v4, which allows users to take advantage of a wide range of plugin-like features with liquidity pools. This paper provides a comprehensive classification and comparative analyses of prominent DEX protocols, namely Uniswap, Curve, and Balancer, in addition to investigating other protocols' noteworthy aspects. The evaluation framework encompasses mechanisms, components, mathematical formulations, and the performance of liquidity pools. The goals are to elucidate the strengths and limitations of different AMM models, highlight emerging concepts in DEX development, outline current challenges, and differentiate optimal models for specific applications. The results and comparative insights can be a reference for web3 developers, blockchain researchers, traders, and regulatory parties.**

*Index Terms*—Decentralized Finance, Decentralized Exchange, Automated Market Makers, Liquidity pool, Comparative analysis.

## 1. Introduction

With the exponential growth and widespread adoption of blockchain technology, various areas and concepts, including online trust, information storage, asset ownership, and financial principles, have undergone substantial infrastructural transformations [1-3]. These advancements have led to the emergence of decentralized finance (DeFi), a significant development that encompasses innovative concepts in financial products. One of the key components within the DeFi landscape is decentralized exchanges (DEXs), providing novel asset-trading mechanisms [4]. DEXs aim to facilitate cryptocurrency transactions in a non-custodial and trustless manner by relying on the execution of smart contracts, in contrast to centralized exchanges (CEXs), which primarily operate using a custodial order book system [5].

To obtain the objective of trustless on-chain asset trading, developers acknowledged the necessity for a new exchange model, as implementing an on-chain order-book-based system proved to be highly inefficient and resource-intensive within blockchain networks [6]. This matter was particularly evident in networks such as Ethereum, where transactions consume higher gas fees. Uniswap emerged as one of the earliest platforms to address this challenge in 2018, attaining remarkable success in trading volume and revenue generation [7]. This platform introduced an automated market maker (AMM) system as an alternative to the traditional order book system. Leveraging Liquidity pools (LPs) and a constant product (CP) formula, Uniswap equilibrated the price of tokens efficiently and accurately to some extent. Subsequently, numerous platforms began to advance the concept of DEX by incorporating additional features and alternative approaches [8].

From a broad perspective, DEXs can be classified into two types: automated market maker (AMM)-based and orderbook-based. In the early stages of DEXs on Layer 1 blockchains, AMMs with a so-called constant product function (CP-AMMs) were a reasonable and creative approach for meeting users' demands adequately. However, further needs,

[1] Corresponding Author
Email addresses: maasef@ut.ac.ir (M.A. Asef), smhosseini@yazd.ac.ir (S.M. Hosseini Bamakan)

more robust tools, and better scalability solutions, such as layer two networks, arose, and a wider range of DEX architectures became possible[9]. A case in point can be DYDX, an orderbook-based DEX with a massive trading volume that has implemented practical features, including zero-gas-fee trading and perpetual markets [10].

Aside from advancements in non-AMM platforms, AMM-based DEXs have also seen game-changing innovations. Both disruptive and market-dominant AMM-based DEXs strive to develop innovative algorithms to achieve better efficiency and higher revenue for liquidity providers and traders [11]. For example, Uniswap, the current predominant DEX of the Web3 space, recently launched its fourth version with many practical upgrades, such as hooks [12]. As another example, Balancer provides a wide range of LP types for various market needs, such as custom-weighted pools [13].

The motivation behind this study stems from the need for a comprehensive overview that evaluates the functionality, components, and efficiency of various DEX protocol types and LP models. Although there have been previous studies in this field, most have concentrated on certain DEX types and evaluated a limited range of their components. Regarding remarkable assessments in this area, a comparison conducted by Bartoletti et al., 2022, explores AMMs from a mathematical perspective [14]. Also, the SOK by Xu et al., 2023 is an outstanding study, providing a detailed assessment of AMM-based DEXs, considering their various dimensions, such as performance and security [15].

In this paper, we aim to provide a comprehensive classification and review of LPs, evaluate the key components of DEXs, including oracles and trading features, and distinguish DEXs in a comparison framework. The paper makes several contributions, which are outlined as follows:

— Reviewed the evolution of cryptocurrency exchanges from outdated CEXs to the latest cutting-edge features of DEXs based on several perspectives, such as the level of decentralization and underlying logic
— Demonstrated the role and significance of DEXs and AMMs within the realm of the DeFi. To do so, we highlighted the integrations of DEXs with other areas of the web3 landscape and examined different versions and components of DEXs in detail.
— Provided a comprehensive analysis of AMMs and liquidity pools in leading DEXs, employing a comparative approach for distinguishing attributes of protocols
— Explained current research challenges and proposed future directions for advancing DeFi protocols

The reminding sections are as follows: Section 2 overviews DEX protocols and outlines their fundamental characteristics. This section presents a classification and comparative analysis of various cryptocurrency exchange types from decentralization level and infrastructure perspectives. Next, section 3 delves into the foundational elements of DEX's structure, encompassing actors, assets, and pivotal modules. Then after, section 4 analyzes the underlying framework, versions, important features, and mathematical logic of three major AMM-based DEXs: Uniswap, Curve, and Balancer. Additionally, this section mentions several other protocols with notable features. Section 5 compares the LP and AMM types employed by the DEXs mentioned earlier in multiple categories. Finally, in section 6, we concluded the paper by summarizing findings and suggesting future works.

## 2. Background

### 2.1 Historical trajectory of cryptocurrency exchanges

The emergence of Bitcoin in 2009 led to the development of cryptocurrency exchanges, platforms that leverage cutting-edge technology to facilitate the buying, selling, and trading of digital currencies [16, 17]. Initially, these exchanges were focused on Bitcoin transactions, but they have since expanded to include Ethereum and its associated tokens [18, 19]. Technological advancements and regulation changes have played a pivotal role in transforming these exchanges, enhancing their accessibility, efficiency, and global reach. Integrating various asset classes and the proliferation of digital platforms have further broadened investment opportunities for institutions and individuals [20].

The different types of cryptocurrency exchanges can be categorized according to their level of control over user funds. The level of centralization and decentralization varies in different exchange models. The first group, CEXs, represents the traditional form of cryptocurrency exchanges, in which a centralized entity exercises control over all aspects of trading, including order matching, fund custody, and transaction settlement. Users deposit their funds into the exchange's wallets, and the exchange acts as an intermediary for all transactions. Prominent examples of CEXs include Coinbase and Binance [21, 22].

Besides, permissioned decentralized exchanges, rendered almost obsolete nowadays, although they are built on top of decentralized infrastructures, require users to obtain permission or fulfill specific criteria before engaging in trading activities. These exchanges often impose stricter Know Your Customer (KYC) regulations and may restrict access based on factors such as geographic location [23]. They also have limited tokens for trading, and token listing involves registering and paying. Permissioned exchanges typically adhere more closely to regulatory frameworks, with Binance DEX as an example of a failed project in this area [24].

Moving to permissionless DEXs, they can have different architectures and system designs. Some DEXs use an off-chain infrastructure in order-matching operations, while others rely on an on-chain infrastructure. Some projects like DYDX ask users to deposit ETH into a smart contract before trading, and the user cannot access the trading tokens on their wallet. Conversely, on the most well-known DEXs using AMM, users have access to their tokens in their wallets and receive the traded tokens instantly on their wallets [25, 26]. Regarding this matter, Table 1 presents an overview of CEX and DEX types from decentralization perspective and details their authorization level with mentioning their limitations.

*Table 1 – Cryptocurrency exchange types from the level of decentralization perspective [21, 27-29].*

| Type | Centralized Exchange (CEX) | Permissioned DEX | DEXs with off-chain operations | Orderbook-based DEX | AMM-based DEX |
|---|---|---|---|---|---|
| Example | Coinbase | Binance DEX (Faild) | EtherDelta (Faild) | DYDX | Uniswap |
| Decentralization limitations | KYC is often required, limited tokens, centralized settlement | KYC is often required, limited tokens, token listing fee is often required | Off-chain order matching and liquidity pools | An initial deposit is locked, no instant access to tokens, restrictions for withdraws can be applied | Oracles that rely on off-chain data are often used |

### *2.2 DEXs in the DeFi landscape*

DeFi comprises a range of financial products built on blockchain technology that seek to establish open, permissionless, and decentralized alternatives to conventional financial intermediaries. By harnessing the capabilities of smart contracts, DeFi aims to redefine the necessity of intermediaries such as banks, brokers, and CEXs. DEX is a critical constituent within the realm of DeFi that facilitates cryptocurrency trading without the involvement of a centralized intermediary. DEXs function through smart contracts that enable users to engage in decentralized asset trading [30].

However, the significance of DEXs within the DeFi domain extends beyond their role as mere trading platforms for traders. As the DeFi ecosystem expands, DEXs have become integral components of numerous platforms requiring direct and indirect interactions. Many tools and services have been designed to enhance the accuracy of DEXs or utilize some functions of DEXs within their services [31].

Various types of decentralized applications (dApps) function by utilizing the capabilities of DEXs, namely lending platforms, algorithmic stablecoins, cross-chain bridges, and DEX aggregators. A case in point is GYD algorithmic stablecoin, which aims to remain stable and pegged to the US Dollar by relying on AMMs and DEXs' features [32]. Another example is yield farming dApps, allowing users to earn rewards by contributing liquidity to DEXs or lending platforms. DEX aggregators are other notable utilizers of DEXs that grant users access to better prices and deeper liquidity for their trades by comparing DEXs. lastly, fundraising, in the form of an Initial DEX Offering (IDO), is another application of some DEXs that enables projects to raise funds by directly selling tokens to the public [33].

Moving to services that DEXs use, some of them are integrated with insurance protocols that offer insurance coverage for smart contracts, safeguarding users against potential risks and vulnerabilities. Another example is governance platforms such as snapshot.org that allow web3 users to participate in the decision-making processes of decentralized organizations, such as DEXs, by voting on proposals and shaping the project's future [34, 35]. Also, decentralized oracles are pivotal components of many DEXs for increasing the accuracy of prices by bringing external data to smart contracts [36].

Table 2 provides a few examples of the relationships between DEXs and other DeFi dApps, listing examples of services a DEX can provide to other dApps to function more effectively, in addition to services a DEX might use from other DeFi platforms to attain better performance [37].

*Table 2- Examples of DeFi services that DEXs utilize, compared to platforms that DEXs enhance [24, 38-40].*

| DEXs Utilize | DEXs Enhance |
|---|---|
| **Oracle Servies** like Chainlink, to ensure the accuracy of prices and other parameters | **Stablecoins** like Gyroscope directly rely on DEXs and many projects interact indirectly |
| **Insurance Platforms** like InsurAce, to cover the on-chain trading risks for users | **Lending protocols** like AAVE operate by relying on DEXs for multiple functions |
| **Governance Protocols** like Snapshot, to drive the application based on users' decisions | **Cross-chain DEX and Bridges** like Jumper use DEXs in their cross-chain transfers |
| **Aggregators** like 1inch, to suggest the lowest prices form other DEXs and find efficient routes | **Aggregators** like 1inch checkout the prices of different DEXs and suggest the best price |

### 2.3 DEX classification

As discussed in Section 2.2, multiple DEX models exist, and each can be classified into sub-models. From the implementation infrastructure perspective, these models can generally be grouped in two: First, DEXs that utilize AMMs and LPs, and second, DEXs that do not rely on liquidity providers and instead adopt an order book model. Aggregators serve a different function from these categories because their primary role is to determine the most favorable prices and trading routes among AMM-based DEXs rather than swapping assets using their own liquidity pools. Additionally, there are hybrid platforms that combine elements of both AMM-based DEXs and orderbook-based DEXs [41-43].

Regarding LPs, DEXs leverage various types of AMM systems that are tailored to different trading pairs. Each pool type caters to specific trading conditions, including price ranges, asset weights, and asset types. These pools utilize distinct formulas, resulting in varying levels of complexity, transaction execution speed (Shang et al., 2023), and revenue for liquidity providers. Notably, a number of DEXs, such as Sushiswap, offer additional incentives in the form of liquidity, providing extra rewards to attract a larger user base [44, 45]. Beyond simple spot swaps, some DEXs provide limit order trades, which operate through off-chain order databases and are filled through arbitrage whenever the price gap and the associated gas fee present a reasonably profitable opportunity. Fig. 1 illustrates a classification of DEXs from the implementation method perspective and pinpoints the variety of LPs and and mathematical logics in the context of AMM-based DEXs [46, 47].

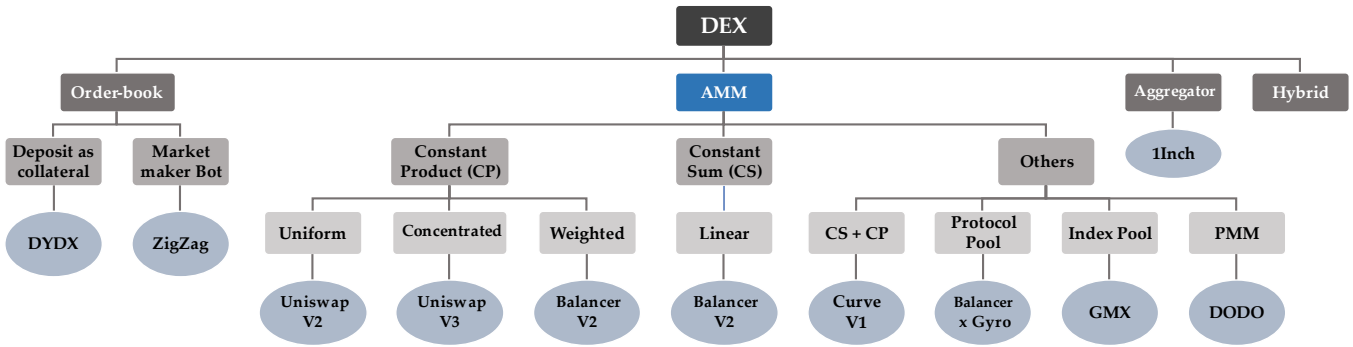

Figure 1: Classification of DEXs from infrastructure perspective and subclasses of AMMs based on conservation functions and LP models.

### 2.4 Automated Market Maker (AMM)

AMM, which stands for Automated Market Maker, is a type of DEX mechanism that enables crypto trading without a traditional order book system. Instead of matching buying and selling orders, AMMs utilize liquidity pools, smart contracts that facilitate swapping assets using an amount of liquidity that is already deposited by liquidity providers. The notable feature is that prices are determined algorithmically based on the ratio of assets in the pool, using mathematical formulas like the constant product. There is a significant variety of logic behind AMMs. [48].

There are various actors, assets, and components within an AMM-based blockchain. These elements can be customized for specific use cases. For instance, protocols may consider an additional incentive to absorb more liquidity providers, adjust the math formula behind the AMM to perform better in specific conditions such as trading stablecoins, or even implement the protocol so that users can add their desired plugins to their AMM [2, 8, 49].

## 3. Components of AMM-based DEXs

To initiate a new LP, a primary liquidity provider supplies an initial amount of at least two crypto assets into a pool. After that, other liquidity providers can increase the LP reserve by contributing additional assets. The providers receive pool shares in proportion to their supply. As an incentive for this contribution, liquidity providers receive transaction fees from exchange users. By redeeming pool shares, liquidity providers can withdraw funds from the pool. The primary liquidity provider should initiate the pool with a ratio of token amounts due to current market prices. In some cases, liquidity providers are referred to as liquidity miners because they also receive newly generated protocol tokens as a reward, in addition to transaction fees [50-52]. The rest of this section defines the main elements of AMMs including actors, assets, and pivotal modules, as illustrated in Fig. 2

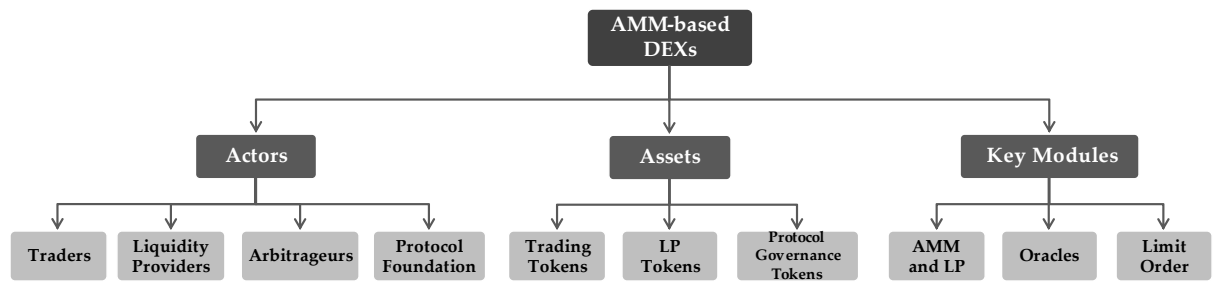

*Figure 2 - Actors, assets, and pivotal modules of AMM-based DEXs [11, 15, 35].*

### 3.1 Actors

#### 3.1.1 Protocol foundation

The protocol foundation comprises founders, the development team, and users with voting rights committed to enhancing the platform. Planning a practical scheme for distributing incentives and revenue among foundation members is necessary to reach further enhancement and success for the platforms [5].

#### 3.1.2 Traders

Traders refer to platform users who wish to exchange their tokens for another token in spot swap platforms or trade crypto assets in derivative platforms. These actors specify the details of their request and submit it through an on-chain or off-chain transaction due to the platform type [6, 53].

#### 3.1.3 Arbitrageurs

Arbitrageurs are a segment of traders and, in most cases, automatic bots who aim to benefit from finding price mismatches in different LPs on DEXs and even the price gap between DEXs and CEXs. Their action stabilizes the price of cryptocurrencies on different platforms [54].

#### 3.1.4 Liquidity providers

Liquidity providers are users who supply crypto assets to the DEX's LPs. This enables traders to execute cryptocurrency exchanges using funds supplied on the LP without relying on centralized intermediaries.

Providing liquidity is incentivized by distributing a portion of the trading fees generated by the DEX [55, 56].

### 3.2 Assets

#### 3.2.1 Trading tokens

Trading tokens are supplied to LPs by liquidity provider users and traded by users. These tokens include a wide range of fungible crypto assets, including native tokens of networks, stablecoins, governance tokens, wrapped tokens, and many other types minted through a particular, standard smart contract. These tokens can be paired together and initiate an LP, e.g., WBTC-USDT, with various pool structure implementations explained in DEXs' components section [57].

#### 3.2.2 LP tokens

LP tokens, short for Liquidity Pool tokens, represent suppliers' shares in LPs. These tokens distribute to liquidity providers based on their supplied amounts in an LP. These shares are the basis of distributing trading fees whenever a trade is executed by the pool. Also, these shares can be redeemed in exchanged for the originally supplied tokens [57, 58].

*3.2.3    Protocol governance tokens*

Protocol tokens represent voting rights in the DEX protocol's decision-making process. These tokens can also be distributed by DEXs as incentives for liquidity providers or airdrop for active traders. Aside from this use cases, protocol tokens can be used for investment purposes and be traded at various kinds of exchanging platforms [20, 57].

## 3.3  Pivotal Modules

*3.3.1    LP and AMM*

LPs are smart contracts that typically hold reserves of two or more tokens, enabling users to trade directly against these reserves. They operate by relying on AMMs based on mathematical formulas to determine prices. They can be implemented with a variety of implementation models. In most DEXs, each LP is a smart contract in which the reserves and trading logic are declared. On the other hand, some DEXs, such as Balancer, have separated the reserves pool and the swapping logic, having a single vault that can operate with multiple custom logics.

Many parameters can be adjusted in LPs, namely incentive allocation calculations, the number of tokens that form the LP, the weight of tokens within the pool, and the mathematical formula that defines the curve of LPs' conservation function. For instance, while in most DEXs, users can initiate pools by pairing any tokens, in a few DEXs, users must pair tokens with a base asset. This base asset is called BNT in Bancor, and it was ETH on Uniswap V1 [46, 59, 60].

*3.3.2    Oracles*

Oracles play a crucial role in AMMs, and various types of oracles are used in DEXs. Two primary categories of oracles are commonly employed in AMMs to provide precise price information on DEXs: off-chain oracles that retrieve price data from external sources such as CEXs or price feeds and on-chain oracles that obtain price information from smart contracts on the blockchain. Off-chain oracles are considered more reliable but may have longer execution latency, while on-chain ones are faster but may be less reliable. The optimal oracle type depends on the platform's infrastructure and specific requirements [61, 62].

*3.3.3    Limit orders system*

Limit orders are considered an extra feature in some AMM-based DEXs and aggregator platforms, while they are commonly used in traditional CEXs. This feature enables users to place a predetermined price for their order, which arbitragers will execute. The execution process can either occur on-chain or off-chain. Most protocols utilize a fee-free method for placing limit orders, as these orders interact with the order-book APIs [63]. Details of DEXs with limit order feature are mentioned in section 4.2

## 4. Market-dominant AMM-based DEXs

In this section, we examine Uniswap, Curve, and Balancer, three preeminent DEXs in terms of Total Value Locked (TVL), trading volume, being multi-chain, and having exclusive features, with a primary focus on spot trades. This section encompasses the evolution of each DEX's history and versions, providing a background in features and attributes that will be compared and analyzed in section 7. Regarding the reasons for this selection, DEXs such as Aerodrome, Raydium, and LFJ have higher figures than Curve and Balancer in some cases. However, they are the primary DEXs of Base, Solana, and Avalanche networks only and are not generally considered dominant DEXs with unique features and a notable evolving history. Also, the PancakeSwap, which is currently multi-chain and has a high amount of TVL and volume, is generally known as a fork of Uniswap on the BSC network. However, many other DEXs with noteworthy features and approaches will be reviewed in sections 5 and 6 [31, 64-66].

Table 3 indicates the TVL and trading volume of Uniswap, Curve, and Balancer, in addition to their market share. As it is evident, Uniswap is the predominant DEX of the crypto market with a considerable distance from other competitors.

*Table 3 – TVL, volume, and market share of three leading DEXs: Uniswap, Curve, and Balancer [64, 65, 67].*

|  | **Uniswap** (25 chains) | **Curve** (8 chains) | **Balancer** (9 chains) |
| --- | --- | --- | --- |
| **TVL** (on 1 Oct 2024) | ~ $4.3B | ~ $1.7B | ~ $0.7B |
| **Market share by TVL** | ~ 24% | ~ 10% | ~ 4% |
| **Volume** (From 1 Jan to 1 Oct 2024) | ~ $571B | ~ $69B | ~ $23B |
| **Market share by volume** | ~ 51% | ~ 6% | ~ 2% |
| **Launched** | Nov 2018 | Jan 2020 | Feb 2020 |

### *4.1 Uniswap*

#### *4.1.1 Overview*

Uniswap has been the most popular, profitable, and used AMM-based DEX science the emergence of DEXs. It is well-known as the first successful project that implemented the famous x*y=k formula, proposed in an article by Vitalik Butrin, the co-founder of Ethereum. This DEX has introduced three up and running versions and a forthcoming version so far [68, 69].

#### *4.1.2 Uniswap V1/V2*

Uniswap v1, the initial version of the protocol introduced in 2018, employed a constant product (CP) conservation function to calculate prices within the LPs. This version only supports ETH/ERC20 swapping pairs. Uniswap v2, launched in May 2020, introduced several novel features. Firstly, ERC20/ERC20 pairs are enabled instead of ERC20/ETH pairs on Uniswap V1, where ETH is being used as the base asset for pairs. Also, a price oracle enabled external contracts to estimate time-weighted average prices over specific time intervals. Additionally, the "flash swaps" feature allowed traders to acquire assets and utilize them elsewhere before settling the payment in a transaction, and a protocol fee was added to the platform. Alongside the new features, Uniswap v2 underwent contract re-architecture to reduce the potential for attacks [7, 70, 71].

Uniswap v1 and v2 both utilize one of the initial implementations of a CP AMM formulas Eq. 1 for LPs encompassing two assets [72].

$$x \times y = k \tag{1}$$

In this formula, x and y represent the number of two tokens with reserves in an LP on a DEX that can be exchanged with each other. The exchange price is calculated due to the ratio of x and y so that the product x × y remains conserved. In fact, by selling Δx tokens, the trader receives Δy tokens [71, 72]. At the time of swapping, reserves change based on Eq. 2.

$$x' = x + \Delta x = (1+\alpha)x = \frac{1}{1-\beta}x$$
$$y' = y - \Delta y = \frac{1}{1+\alpha}y = (1-\beta)y \quad (2)$$

### 4.1.3 Uniswap V3

Uniswap v3, Released in May 2021, focuses on enhancing capital efficiency for liquidity providers, optimizing the accuracy of the price oracle, making the fee structure more adjustable, and introducing a liquidity oracle. Regarding the price Oracle, V3 accommodates user queries for recent price accumulator values, obviating the necessity of explicitly checkpointing the accumulator value at the precise beginning and end of the period for which a time-weighted average price is being computed. In liquidity oracle, the contracts expose a time-weighted average liquidity oracle, furnishing valuable insights into liquidity trends. Also, the previous fixed swap fee of 0.30% was replaced with a fee tier system. Each pool, which could encompass multiple asset pairs, has its fee tier set upon initialization, with initial options including 0.05%, 0.30%, and 1% [70, 73, 74].

Beside these changes, the foundational concept behind Uniswap v3 is the notion of concentrated liquidity, which entails liquidity confined within a specific price range. In Uniswap V1 and V2, liquidity was uniformly distributed along the curve described by Eq. 1 to offer liquidity across the entire price range (0 to ∞). Although this approach is straightforward to implement and allows for efficient liquidity aggregation, it also means that a substantial portion of the assets held within an LP may remain untouched. However, concentrated liquidity within a finite range leads to more liquidity engagement, which results in more trading fees for liquidity providers [75-79].

In V3, the price range is divided into "ticks," representing discrete price points. The formula is applied between these ticks. Instead of using actual token amounts, V3 uses "virtual" reserves calculated based on the current price and the liquidity within the active range. The key innovation is the liquidity (L) parameter, which represents the amount of liquidity available at a given price range. In this version, x and y are as Eq. 3, Where L is the liquidity amount, P is the current price, $P_b$ is the upper price bound of the range, and $P_a$ is the lower price bound of the range[75, 80, 81].

$$x = L(\sqrt{P_b} - \sqrt{P}), \quad y = L\left(\frac{1}{\sqrt{P_a}} - \frac{1}{\sqrt{P}}\right) \quad (3)$$
$$\left(x + \frac{L}{\sqrt{P_b}}\right)\left(y + L\sqrt{P_a}\right) = L^2 = k$$

As mentioned in Eq. 3, the value of L, which equals the square root of k, can be utilized to measure the liquidity provided. The curve in figure 3 characterizes the position's actual reserves.

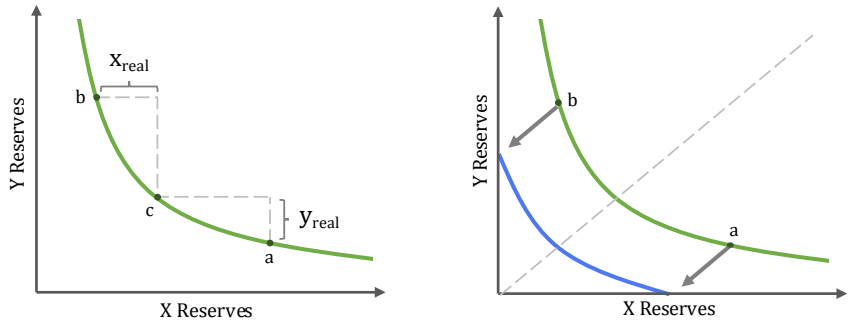

*Figure 3 – The visualization of concepts of virtual reserves (green line) and real reserves (blue line) in Uniswap v3 [75].*

### 4.1.4 Uniswap V4

Uniswap v4, currently in its alpha phase, has introduced innovative concepts across various domains. A primary advancement is the introduction of hooks, plugin-like add-ons that enhance programming customized features on top of liquidity pools. In addition, Uniswap v4 enhances gas efficiency through singleton implementation of LPs' smart contracts. Doing so, creating a pool is a state update instead of the deployment of a new contract and swapping through multiple pools does not requires transferring tokens for intermediate pools. Doing so, initiating a pool is a state update instead of deploying a new smart contract [70, 82].

Another practical feature is flash accounting; using EIP-1153, balance changes are recorded in transient storage efficiently, enabling users to pay only the final balance change without resolving intermediate balance changes. Also, the feature of dynamic fees lets LPs modify their fees. The frequency of fee updates is also flexible, and they can occur on every swap, every block, or on a custom schedule. Finally, this version supports native token assets (ETH) without wrapping/unwrapping the native token to WETH, which is required in previous versions [82].

In Uniswap v4, any individual can create a new pool featuring a specified hook that can be executed before or after predefined pool actions. Hook is as a vehicle used to implement features that were previously integrated within the protocol, such as oracles, while also enabling the inclusion of novel functionalities that would have otherwise required independent implementations of the protocol. Example projects that have been integrated Uniswap hooks such as Brokkr.finance are explained in section 5. As an example of utilizing hooks, Figure 4 demonstrates swapping process, where a before and after swap hooks can called withing the flow [12, 83].

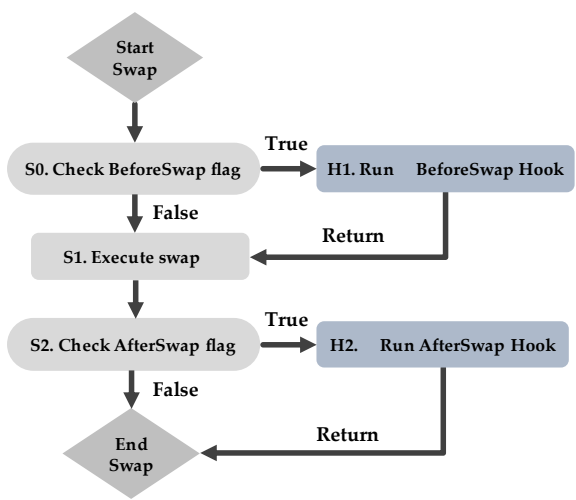

*Figure 4 - The flow of using hooks within a swapping process on Uniswap V4 [12].*

### 4.2 Balancer

#### 4.2.1 Overview

The Balancer is a multichain decentralized AMM protocol representing a flexible building block for programmable liquidity. By separating the AMM curve logic and math from the core swapping functionality, Balancer becomes an extensible AMM that can incorporate any number of swap curves and pool types. Balancer v1 was founded in 2019, and Balancer v2 was launched in March 2021. The new version of the platform offered several improvements over v1, including a wider variety of pool types, the ability for pool creators to set their own fees, and more liquidity incentives, such as boosted rewards and liquidity bootstrapping [67, 84]

#### 4.2.2 The Vault

The underlying structure of Balancer is distinct from many other DEXs. The Vault, a comprehensive smart contract, is the core of all liquidity pools and is in charge of managing all tokens in each pool type and handling operations, such as joining and exiting LPs. Adaptable to systems that fulfill required standards, the Vault is agnostic to the mathematical logic of the LP. By doing so, developers can implement their own customized swapping schemes and plug them into the liquidity of the Balancer protocol. Regarding this concept, figure 5 illustrates the architecture and components of a single vault that holds and manages tokens along with functions that can be handled [84, 85].

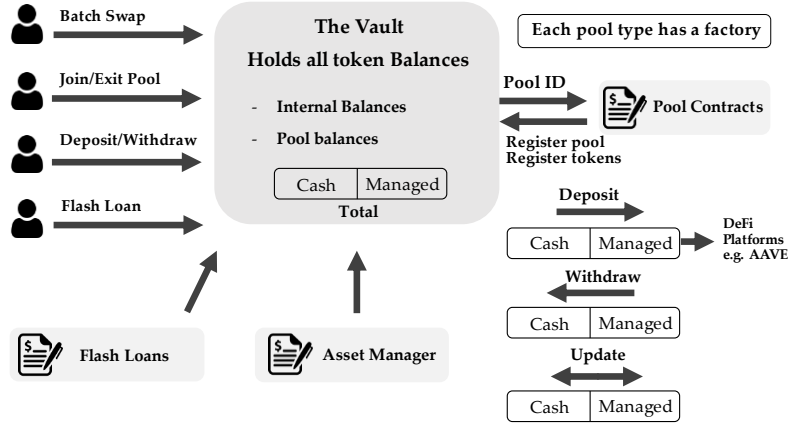

*Figure 5 - The architecture of a vault on Balancer V2, including critical components and applicable functions [86].*

### 4.2.3 Pool types

#### 4.2.3.1 Weighted pool

Weighted Pools represent a generalized version of the constant product formula used in Uniswap. These pools allow users to initiate LPs with more than two tokens and custom weightings. For example, a pool with 70/30 or 50/30/20 weightings, in contrast to many DEXs that only offer a 50/50 ratio. The math formula for a weighted pool is as Eq. 4, where t represents the tokens in the pool, $B_t$ stands for the balance of tokens, and $W_t$ shows the weight of tokens, such that the summation of all weights is equal to 1 [84].

$$V = \prod_t B_t^{W_t} \tag{4}$$

#### 4.2.3.2 Composable stable pool

Composable Stable Pools are designed for exchanging stablecoins, e.g., USDT/USDC, or correlated Tokens, e.g., wstETH/wETH, and pegged assets in general. This pool uses the stable math formula based on the famous StableSwap by Curve.fi to improve capital efficiency for like-kind swaps without a marked price impact. The math formula for a composable stable pool is as Eq. 5. In this formula, n is the number of tokens, $X_i$ is the balance of tokens, and A is the amplification parameter [84, 87].

$$n^n A \sum x_i + D = ADn^n + \frac{D^{n+1}}{n^n \prod x_i} \tag{5}$$

#### 4.2.3.3 Linear pool

Linear Pools are best suited for exchanging tokens with their wrapped or yield-bearing replications where the exchanging ratio is already determined. A case in point can be the ETH/aETH pair (Aave-wrapped ETH). These pools utilize a so-called fee/reward mechanism to incentivize arbitrageurs to conserve an exact ratio between the two tokens. In detail, by leaving the target ratio, they pay a fee, and by returning to the range, they earn a reward.

The mathematical formula behind this pool is based on the constant sum invariant. This formula is presented in Eq. 6, where t represents tokens in the LP, $B_t$ shows the balance of the tokens, and $R_t$ stands for the target exchange rate. R is equal to 1 for the primary unwrapped token [88].

$$I = \sum_t (B_t * R_t) \tag{6}$$

#### 4.2.3.4 Liquidity bootstrapping pool (LBPs)

Liquidity Bootstrapping Pools (LBPs) are capable of changing the weight of tokens dynamically. These pools rely on the weighted math that was described earlier with time-dependent weights. The LP owner can specify the initial and ending weights, adjust times, and even stop the LP's operations. In these pools, only a single address can join the pool, which is eventually their LP owner. In LBPs, a high initial price should be set to allow the changing pool weights to lower the price gradually until the market becomes equilibrated. In contrast to many AMM schemes, LBP users should wait until the price decreases to a fair level [84, 89].

*4.2.3.5 Managed pool*

Managed pools enable the implementation of advanced portfolio management strategies and provide a high level of control and customizability over LPs. Utilizing weighted math, these pools can contain up to 50 tokens, making them a suitable DeFi product for implementing on-chain fund management techniques. Also, these pools are equipped with weight-shifting and time-based mechanisms similar to those provided by LBPs. [84].

*4.2.3.6 Protocol pool*

In contrast to other pool categories mentioned previously as particular LPs, Protocol Pools are defined as DeFi products built on top of the Balancer protocol infrastructure. As Balancer has separated the pool logic from the accounting logic, projects can develop their desired AMM logic as a custom pool upon Balancer Vault or, in other words, Balancer's programmable liquidity. Gyroscope protocol (Gyro.finance) is a well-known example of the protocol pool utilizer. They have developed a robust stablecoin project on top of Balancer infrastructure by developing their customized concentrated liquidity pools (CLP) in several variations and using LP capabilities as the backbone of their stablecoin, GYD. [33, 67, 84].

## 4.3 Curve

### 4.3.1 Overview

Curve is a leading AMM-based DEX that with a primary focus on efficient trading of stablecoins alongside other types of tokens. Additionally, Curve has introduced their own stablecoin, crvUSD, and Curve Lending, which both have a distinct liquidation mechanism called LLAMMA [90].

### 4.3.2 StableSwap pools

The StableSwap, also known as Curve V1, combines attributes of both constant product and constant sum invariants to structure an ideal mechanism for swapping the pegged assets. The stable swap includes three subclasses of plain pools, lending pools, and meta pools, each customized for specific purposes. The mathematical formula of StableSwap is mentioned in Eq. 7, where D represents invariant, xi denotes the initial balance of each token, n is the number of assets in a pool, and A is the amplification coefficient selected by the protocol team. In this formula, as the variable A tends to 0, the outcome becomes similar to a constant-product mechanism, and when it tends to $+\infty$, the result would be similar to a constant-sum function with an target rate of 1 [91].

$$An^n \sum x_i + D = ADn^n + \frac{D^{n+1}}{n^n \prod x_i} \qquad (7)$$

*4.3.2.1 StableSwap sub-classes*

- **Plain pool** is a straightforward implementation of StableSwap logic. A primary point of this pool is the fact that the LP smart contract is the holder of all the supplied tokens at all times.
- **Lending pools** are equipped with lending capability so that they can lend out tokens on a number of integrated dApps, such as AAVE, Compound, or Yearn. In contrast with plain pools, where users hold the underlying tokens, in lending pools, users keep a wrapped representation of assets. As an example, a Curve-Compound LP includes cUSDC and cDAI as wrapped tokens, whereas USDC and DAI are supplied on Compound.finance as underlying tokens. In this scenario, the liquidity providers earn Compound protocol rewards plus regular swapping fees [92-95].
- **Meta pools** pair stablecoins, such as GUSD, with LP tokens from other base pools, such as 3CRV. Liquidity providers in these pools receive Metapool LP tokens, which can be used for extra incentives if staked.

- **Stableswap NG (New Generation) pools** utilize similar StableSwap math logic but benefit from two new major features of dynamic fees and a price oracle. [91, 92]

*4.3.3 CryptoSwap pools*

CryptoSwap pools aim to facilitate swapping tokens that are not necessarily pegged to a particular asset by utilizing an internal oracle and a dynamic peg mechanism. The AMM concentrates liquidity within a range close to the market price that is calculated by the internal oracle. The goal is to maximize the efficiency and utilization of liquidity and, as a result, increase earnings for the liquidity provider, similar to what Uniswap V3 does. Another feature of this version is a dynamic fee system that functions according to market conditions [35, 94].

There are several sub-classes of CryproSwap: Cryptoswap, the genesis implementation with two tokens; Tricrypto, the genesis implementation with three tokens; Twocrypto-NG and Tricrypto-NG, the improved versions of genesis pools Tricrypto. Crypto-NG pools are an optimized version of the cryptoswap genesis pool with an improved auto-rebalancing feature and utilizing the price oracle that is also used in stable swap NG [96].

Eq. 8 mentions the logic behind crypto swap pools, where D represents invariant, n shows the number of tokens in the LP, $x_i$ is the balance of the token, A stands for the amplification coefficient, and $\gamma$ denotes the amount between two possible curves illustrated with dashed lines in Figure 6. This figure also depicts how the crypto swap invariant behaves as a combination of constant-product and stable swap invariants [93, 97].

$$KD^{N-1}\sum x_i + \prod x_i = KD^N + \left(\frac{D}{N}\right)^N$$
$$K_0 = \frac{\prod x_i N^N}{D^N}, K = AK_0 \frac{\gamma^2}{(\gamma+1-K_0)^2}$$
(8)

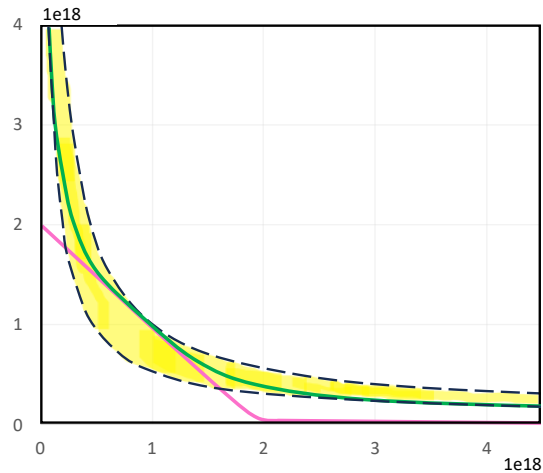

*Figure 6 – Crypto swap invariant (green) vs. to stable swap (pink) vs. constant product (dashed line) which is being used in Uniswap [97].*

## 5. Other DEXs with notable features

### 5.1 Kyber and Matcha: DEXs with limit order feature

In DEXs such as Kyber, limit orders are stored on the blockchain. Conversely, protocols like Matcha took a different approach by storing orders off-chain, with only the trade settlement process utilizing the blockchain. Once a limit order has been fulfilled, the transaction settlement takes place on the blockchain, wherein the individual placing the order is responsible for covering the associated network gas fee [98].

Additionally, users can specify an expiration period for their limit order and retain the option to cancel it. Cancelling an order prior to its designated expiration time necessitates a nominal gas payment to cover the transaction fee for the cancellation process. However, orders assigned an expiration time will be automatically canceled without incurring any cost. Regarding the process behind this module, Figure 7 illustrates the process of a limit order on Matcha DEX aggregator. This figure shows the actors and functions in on-chain and off-chain layers. It is notable that not only Matcha, but many other DEXs and aggregators are using such process [39, 63, 99].

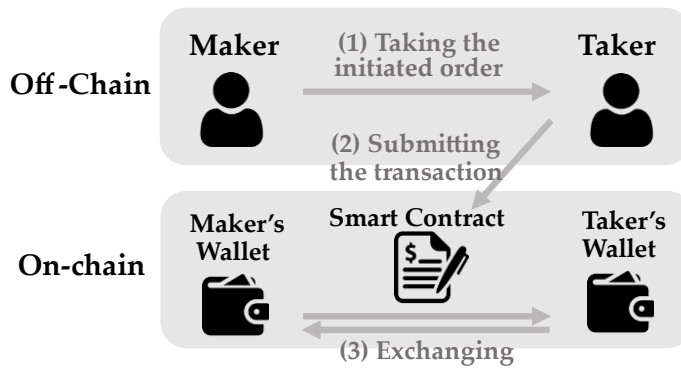

*Figure 7 – Actors and function of Matcha limit order process in Off-chain and On-chain environments [63].*

### 5.2 DODO: Proactive market maker (PMM)

DODO is a DEX powered by the PMM algorithm. PMM can be described as an optimized and adapted on-chain generalization of order book trading, offering higher capital efficiency and lower slippage than some AMMs. PMM relies on external price oracles to determine the "real" market price of assets. Liquidity in PMM is divided into two parts: the Base Token, which has more liquidity, and the Quote Token, which has less liquidity.

In DODO, new assets can be issued at the lowest cost, and highly liquid capital pools can be formed through crowdpooling without restricting the initial price or the ratio of funds and tokens provided. DODO is somewhat a DEX combined with an aggregator with its own LPs that support trading between any two tokens in the market.

The actual formula behind PMM's price determination is considerably more complicated than that of many AMMs. Eq. 9 is a simplified version of it: $P_m$ is the price offered by the PMM, $P_o$ is the price from the external oracle, and k is a parameter that determines the price range [15, 100, 101].

$$P_m = P_o (1 \pm k) \qquad (9)$$

### 5.3 GMX: An AMM-based derivative DEX with a so-called GLP pool

GMX is a well-known protocol in the perpetual trading DEXs sector. This app, which is deployed on top of Arbitrum and Avalanche layer-two networks, provides spot trading as well, but with a different infrastructure than many DEXs. To facilitate trading, GMX utilizes a multi-asset, customized LP that incentivizes liquidity provision by distributing fees from market-making, swapping, and leveraged positions plus GMX exclusive rewards. The price determination functions by relying on Chainlink Oracles, which obtain prices by inspecting various legit sources. [102]

GLP, the customized liquidity pool of GMX V1, represents an index of tokens utilized for spot and perpetual trades.

GLP which is the LP token of this pol can be minted using various tokens that are called index assets and are redeemable for any index asset as well. The cost associated with these operations is determined based on the ratio of the value of all assets in the index divided by the supply of GLP [102].

Figure 8 depicts the architecture of the GMX platform, highlights the role of GLP and shows the interactions of the GLP with traders, trading operations, liquidity providers, and price oracles.

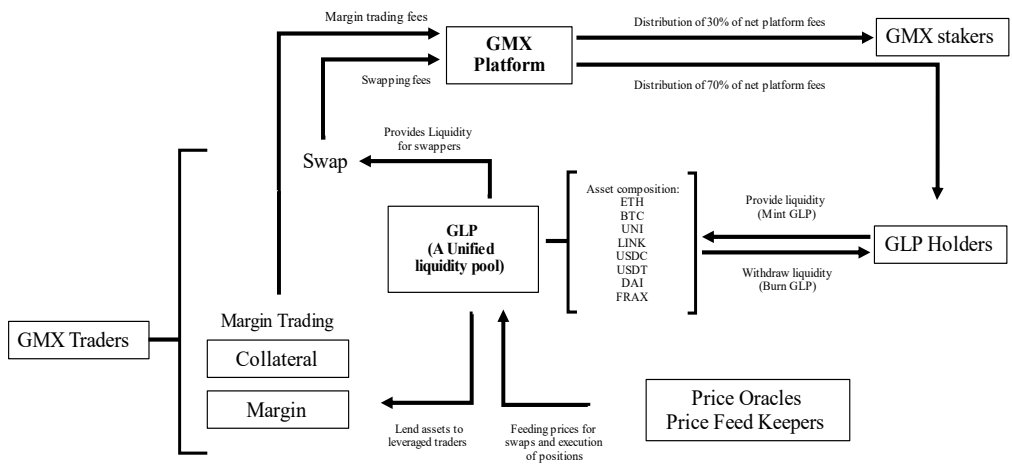

*Figure 8 – The architecture of GMX protocol V1 and the role of GLP, traders, liquidity providers, and oracles [103].*

GMX V2 comprises GLV (GMX Liquidity Vault) and GM (GMX Market) pools. GLV pools support multiple markets and automatically shift liquidity between markets due to the utilization level and recommendations from external sources. Conversely, GM pools are specific to individual markets; users can shift GM tokens between pools, causing price impact but no buy/sell fees. Regarding similarities, both models use long and short tokens to back positions. The pricing of GM and GLV tokens is determined according to the prices of long and short tokens as well as traders' open position PnL. The trading fees increase pool tokens' prices, and pools aim to maintain a balance between long and short tokens. In fact, the system incentivizes pool rebalancing through price impacts when token values change [102].

## 5.4 Brokkr: DEXs integrated with Uniswap V4 hooks

Brokkr Finance is an asset and liquidity management platform that attempts to implement advanced trading features using Uniswap V4 hooks. These hooks function as plugins for LPs, expanding their capabilities beyond traditional liquidity management and trading [104]. Some of the hooks that have been developed as proof-of-concept are as follows:

- Locking Liquidity & Incentives Hook allows users to lock liquidity in the pool and receive rewards. It can function on full-range LPs based on V2 pools with customizable tokens or NFT rewards.
- Dynamic Fees by Volume Hook can adjust swap fees based on trading volume, functioning as a proxy for market volatility. This hook can provide more earnings in high-volume periods during volatile markets.
- Liquidity Management Hook automates position management for liquidity providers, a critical matter for LPs such as Uniswap V3 concentrated pool. This hook can lead to higher profits for liquidity providers with methods such as position rebalancing [83].

## 6. Comparison Framework

In this section, according to the literature review, we considered key features of DEXs and compared the specifications of LPs on AMM-based DEX in three sub-sections. First, we compare the different versions, LP models, and features inside each protocol discussed in section 4. Next, we select LPs with similar functionalities and group them. These groups include full-range LPs, concentrated LPs, stablecoin LPs, and Vaults. Finally, we compare the LP categories to distinguish the differences more vividly. Figure 9 illustrates the framework of these comparisons visually. Consider that details of different specifications, such as mathematical formulas, are already explained in the protocol-specific descriptions in section 4 of the article.

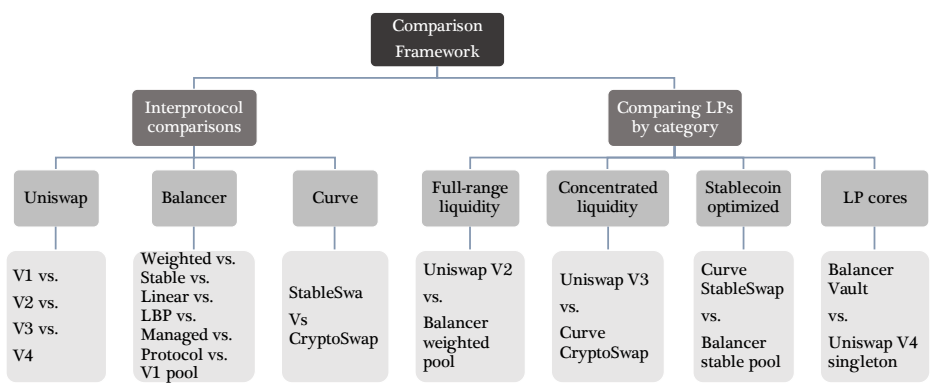

*Figure 9 – Sub-sections of the comparison framework*

### 6.1 Interprotocol Comparisons

In the following, we will compare specifications and market statistics of different versions and variants of Uniswap, Balancer, and Curve DEX protocols.

### 6.1.1 Uniswap

Table 4 compares four versions of the Uniswap protocol from several aspects. It mentions AMM specifications, including conservation function, LP type, mathematical formula, and important features. Also, it details the state of each version in the crypto market based on total value locked (TVL), number of existing pools, and accumulated number of traders and trading volume from 1 Jan 2024 until 1 Oct 2024.

*Table 4 – Comparing of four Uniswap versions based on features, LPs, logic, and market share [12, 64, 65, 67, 71, 75, 105].*

| Version | Uniswap V1 | Uniswap V2 | Uniswap V3 | Uniswap V4 |
|---|---|---|---|---|
| Launched | Nov 2018 | May 2020 | May 2021 | Q4 2024 |
| Function | CP | CP | CP | CP |
| LP type | Uniform | Uniform | Concentrated | Both V1 and V2 pools |
| Invariant formula | $x \times y = k$ | $x \times y = k$ | $(x + \frac{L}{\sqrt{P_b}})(y + L\sqrt{P_a}) = L^2 = k$ | Both V1 and V2 logics |
| Feature | Simple \| 0 to ∞ liquidity range \| WETH as base asset for all LPs | 0 to ∞ liquidity range \| ERC20/ERC20 pairs \| TWAP oracle \| flash swap | Custom liquidity range \| governance token \| enhanced price oracle \| liquidity oracle | Hooks \| singleton contract \| native ETH support \| flash accounting |
| Fees | No fee | 0.3% Flat | Dynamic range | Dynamic and schedulable |
| Num. of LPs | - | ~ 27k | ~ 375k | - |
| TVL | ~ $7M | ~ $1.6B | ~$2.9B | - |
| Traders (2024) | > 10k | ~ 7.8M | ~ 24.5M | - |
| Volume (2024) | ~ $18M | ~ $571B | ~ $165B | - |
| TVL share | ~ 0% | ~ 35% | ~ 65% | - |
| Traders share | ~ 0% | ~ 24% | ~ 76% | - |
| Volume share | ~ 0% | ~ 78% | ~ 22% | - |

The details in Table 4 show that although Uniswap V3 has been successful in absorbing a considerably higher number of traders and LPs, the volume on Uniswap V2 was significantly higher. The reason might be the impossibility of accessing some V2 features on V3. The Uniswap team aims to facilitate the transition to the latest version by allowing users to create both types of V2 and V3 pools in V4. That being said, V1 has become completely obsolete, with almost 0% share in the platforms' volume and TVL.

### 6.1.2 Balancer

Balancer protocol has developed several LP models for different use cases and purposes. Table 5 highlights similarities and differences between the four V2 pools, the protocol pool, and the single V1 pool. Specifications such as purpose, AMM logic from a mathematical perspective, and fee payment method are measured in this comparison. Also, the market state of V2 LPs, V1 LP, and Gyro project, the most well-known example of a protocol pool, are mentioned.

*Table 5 – Comparing seven Balancer LPs from several technical and market market performance aspects [13, 33, 84, 106].*

| LP Name | Weighted | Stable | Linear | LBP | Managed | Protocol | V1 Pool |
|---|---|---|---|---|---|---|---|
| Protocol | V2 | V2 | V2 | V2 | V2 | Built on Balancer | V1 |
| Purpose | Customized assets' weights | Suitable for stablecoin pairs | Suitable for pairs with yield bearing assets | Dynamic change of weights | Exclusive configuration for LP owners | LP infrastructure for projects like Gyro.finance | Initial weighted pool |
| Invariant formula | $V = \prod_t B_t^{W_t}$ | $A \cdot n^n \cdot \sum x_i + D = A \cdot D \cdot n^n + \frac{D^{n+1}}{n^n \cdot \prod x_i}$ | $I = \sum_t (B_t * R_t)$ | weighted math | weighted math | Customizable | weighted math |
| Function | CP | CP+CS | CS | CP | CP | Customizable | CP |
| Tokens | Up to 8 | Up to 5 | Up to 5 | Up to 4 | Up to 50 | Customizable | Up to 8 |
| Fee payment | BPT | BPT - Pool tokens | No fee | Revenue sharing | BPT | Customizable | No fee |
| Add/remove tokens | - | - | - | - | Yes | Customizable | - |
| TVL | ~ $378M | ~ $316M | ~ 22k | ~ $1.1M | ~ $16k | Gyro e.g. ~ $52M | V1: ~ $27M |
| All time Volume | V2 LPs all together: ~ $81B | | | | | Gyro e.g. ~ $2.1B | V1: ~ $8B |

According to Table 5, Balancer V3 comprises a few pools with distinct math formulas such as weighted, stable, and linear, whereas there are other LPs that utilize the core logic of weighted math, namely boosted, liquidity bootstrapping, and managed. The provided information shows that V1 has become almost obsolete, and most liquidity is on V2 pools. Also, the noticeable liquidity on the Gyro project indicates that the protocol pool is a robust LP infrastructure for such LP-focused projects.

### 6.1.3 Curve

Curve protocol has had two major releases, stable swap, and crypto swap. In Table 6, the purpose, features, LP variants, and AMM logic behind them are compared. As each version includes several sub-class LPs and separate protocols, we did not find distinguishing the market data of these subclasses a practical analysis.

*Table 6 – Comparing the two releases of Curve protocol based on features, mathematical formulas, and LP variants [90, 92, 96, 97].*

| Version | Stable Swap | Crypto Swap |
|---|---|---|
| Released | Nov 2019 | June 2021 |
| Purpose | Designed for stablecoins and other pegged assets | Handles both stable and volatile assets |
| Features | Concentrates liquidity around the 1:1 price ratio | Distributes liquidity based on asset volatility |
| | Very low slippage for similarly-valued assets | Variable slippage based on asset volatility and trade size |
| Invariant formula | $An^n \sum x_i + D = ADn^n + \frac{D^{n+1}}{n^n \prod x_i}$ | $KD^{N-1}\sum x_i + \prod x_i = KD^N + \left(\frac{D}{N}\right)^N$  $K_0 = \frac{\prod x_i N^N}{D^N}, K = AK_0 \frac{\gamma^2}{(\gamma + 1 - K_0)^2}$ |
| Function | CS + CP | CS + CP |
| LP Variants | Plain pools, Lending pools, Meta pools, StableSwap NG | Tricrypto, Twocrypto, Twocrypto -NG, Tricrypto -NG |

Table 6 displays that both versions use a similar model of conservation function, utilizing both constant product and constant sum at their formula. Also, both releases include several LP subclasses and new generation (NG) upgrades. Regarding differences, stable swap is well-suited for a stablecoin-to-stablecoin swaps and LPs, while the other version can handle a wider range of tokens adequately.

## 6.2 Comparing LPs by category

This section compares LP models across DeFi protocols by categorical use cases. In the first comparison, we investigate LPs renowned for general-purpose swaps. Secondly, we compare LPs focusing on concentrating liquidity within a specific range. Then, a comparison of stablecoin-compatible pools will be assessed. Comparisons are based on several metrics, including the invariant formula, fee structure, rewarding and incentives, potential risk-rewards, and other notable features. Segmenting by pool type and application provides a nuanced perspective on where competitive advantages exist per protocol.

### 6.2.1 Full-range liquidity

In Table 7, we analyzed the famous Uniswap V2 with the Balancer weighted pool, which can be considered a generalized model of constant-product-based AMM used in Uniswap V2. The comparison distinguishes the underlying math, number of tokens and ratio in LPs, details of liquidity distribution within the pools, and other notable attributes of these protocols.

*Table 7 – Comparing Uniswap V2 and Balancer weighted pool, general-purpose LPs with full-range liquidity [13, 70, 71, 84, 107].*

| LP | Uniswap Uniform LP | Balancer Weighted Pool |
|---|---|---|
| **Launched** | May 2020 | May 2021 |
| **Protocol** | Uniswap V2 | Balancer V2 |
| **Formula** | $x \times y = k$ | $\prod_t B_t^{W_t} = k$ |
| **Function** | CP | CP |
| **Tokens** | 50/50 token ratio \| pairs with 2 assets | Customizable token ratio \| 2 to 8 pairing assets |
| **Liquidity** | Uniform across entire price range (0 to ∞) | Uniform but influenced by weights |
| **Slippage** | Higher risk in pools with low liquidity | More mitigable via custom weightings |
| **Fee** | 0.3% constant | customizable |
| **Key Features** | Higher liquidity in protocol \| more simplicity | More customization and control over LP configuration |

The information in Table 7 pinpoints that the weighted pools provide a more customizable configuration for initiating LPs. That being said, the Uniswap V2 protocol contains significantly higher liquidity. Balancer-weighted LP can be a more suitable option for portfolio management as it can have multiple integrations with Balancer protocol exclusive features, while Uniswap is an absorbing option for traders and liquidity providers who look for volatile markets with deep liquidity.

### 6.2.2 Concentrated liquidity

Table 8 indicates differences between the Uniswap V3 concentrated pool and the Curve crypto swap. Although both aim to benefit from concentrating liquidity within a particular range, their specifications differ, including the liquidity concentration method, the number and ratio of tokens within the LP, and performance potentials.

As the table shows, both protocols do not rely on the external oracle, but Curve uses an internal oracle to adjust the liquidity ranges. In Uniswap V3, however, liquidity providers should follow the market and manually adjust their range to maximize their performance. Overall, Uniswap seems more suitable for active liquidity provision, which brings higher risk and rewards, while Curve facilitates passive liquidity provision with adequate returns. It is worth mentioning that Curve users have a chance of earning CRV tokens in some cases.

*Table 8 – Comparing Uniswap V3 with Curve cryptoswap, Concentrated LPs aiming to maximize earnings [75, 84, 107-110].*

| LP | Uniswap concentrated LP | Curve Crypto swap pool |
|---|---|---|
| Protocol | Uniswap V3 | Curve V2 |
| Launched | May 2021 | June 2021 |
| Formula | $(x + \frac{L}{\sqrt{P_b}})(y + L\sqrt{P_a}) = L^2 = k$ | $KD^{N-1}\sum x_i + \prod x_i = KD^N + \left(\frac{D}{N}\right)^N$ <br> $K_0 = \frac{\prod x_i N^N}{D^N}, K = AK_0 \frac{\gamma^2}{(\gamma+1-K_0)^2}$ |
| Oracles | Provides TWAP oracle | Relies on an internal price oracle |
| Concentration | Manually defined by liquidity providers | Automatically adjusted by protocol |
| Tokens | 2 \| 50-50 ratio | 2 or 3 \| custom ratio |
| Fee | Customizable (up to 1%) | Customizable (up to 3%) |
| Efficiency dependency | Depends on choosing a proper liquidity range | Depends on the performance of internal oracle |
| Risk-reward potential | Higher \| Protocol claims up to 4000x greater capital efficiency than Uniswap V2 is possible | Lower as follows an automated pattern |
| Rewarding | Swapping fees for liquidity providers | Fees + CRV in some cases for liquidity providers |

### 6.2.3 Stablecoin-optimized

As stablecoins or pegged tokens are pivotal components of the DeFi ecosystem, several DEXs attempted to develop LPs optimized for these types of crypto asset. The Curve was a pioneering platform that popularized the stable swap invariant formula, and some other DEXs, such as Balancer, used this logic to develop LPs suitable for assets with a common peg. Table 9 compares Curve and Balancer stablecoin-optimized LPs based on unique advantages, gas optimization solutions, and token structure.

*Table 9 – Comparing two stablecoin-optimized LPs, Curve stable swap and Balancer composable stable pool [84, 87, 90-92].*

| LP | Curve Stable swap | Balancer Composable stable pool |
|---|---|---|
| Protocol | Curve V1 | Balancer V2 |
| Formula | $$An^n \sum x_i + D = ADn^n + \frac{D^{n+1}}{n^n \prod x_i}$$ | |
| Fee | Up to 1% customizable fee | customizable fee |
| Incentive | CRV token in some cases | BAL token in some cases |
| Feature | Multiple LP variants based on stableswap \| the pioneer and dominant DEX for stablecoin LPs | Stableswap under Balancer umbrella. E.g. integration with Batchswap, nesting, and pre-minting features |
| Gas | Several LP sub-classes for different types of stablecoins | Several integrated options for gas efficiency |
| Tokens | Up to 8 \| custom ratio | Up to 5 \| custom ratio |

This table shows that although these LPs use similar math logic, they provide distinct, exclusive features due to their protocols. Stableswap users in the Balancer protocol can utilize and integrate with other LP types and features of the Balancer ecosystem, and Curve users can also choose from a variety of LPs based on stable logic based on their needs.

### 6.2.4 LP cores

In the last part of the comparisons section, we compare two structures in the DEX landscape that aim to have a central asset center with LP models linked to it. Balancer was one of the pioneer protocols in implementing such an idea, and now Uniswap aims to implement a relatively similar concept in its fourth version. Table 10 distinguishes the specifications of these two DeFi products based on architecture, key features, customizability options, gas efficiency, and technical differences in calculation and balance processing methods.

*Table 10 – Comparing two structures that function as a collection of LPs, Balancer vault and Uniswap singleton [12, 82, 84, 85, 89]*

| Structure | Balancer Vault | Uniswap Singleton |
|---|---|---|
| Protocol | Balancer V2 | Uniswap V4 |
| Architecture | A central vault for different LP types | Singleton contract for shared infrastructure |
| Key specifications | Separation of LPs' logic and assets \| LPs initiate with deploying separate contracts | Centralized logic for LPs \| LPs initiate with state update not deploying new contracts |
| Customizability | Integration with Balancer LPs such as composable pools \| multi-asset and custom liquidity weighting flexibility | Hooks enable many customizable LP add-ons such as various oracles, and custom liquidity strategies |
| Efficiency | internal balances and batch trading features | efficient inter-pool interactions |
| Gas | Higher gas cost and modularity due to separate LP deployments | Lower gas cost and less modularity |
| Calculation logic | Pool library | Pool contract |
| Balance Processing | Vault contract | Pool manager contract |
| User interactions | Vault contract | Callback contract |

The table indicates many differences in the functionality of these models. Initiating LPs and connecting new LP types to the Balancer vault requires new deployments, whereas, in Uniswap implementation, a singleton contact handles processes and requests. Balancer's exclusive feature is its integration with numerous Balancer LP-related products, making it a decent choice for enthusiasts in multi-asset flexible pools and yield farming. On the other hand, Uniswap offers hooks as a bridge to numerous custom features that can absorb many web3 developers familiar with Uniswap LPs.

## 7. Conclusion

Decentralized Exchanges (DEXs) are remarkable innovations in the Decentralized Finance (DeFi) landscape. Automated Market Maker (AMM) methods and Liquidity Pool (LP) models are pivotal components of DEXs. These components have experienced numerous advances so far and are expected to continue their development toward more secure and efficient solutions.

This article examined the architecture and components of DEX, classified protocols from different perspectives, and highlighted trade-offs between DEXs. Within comparisons, leading DEXs such as Uniswap, Curve, and Balancer were assessed, as were newer protocols with notable and unique features. The results aim to inform diverse audiences ranging from developers to regulators. As a future work it is suggested to investigate the possible opportunities and threats of utilizing concepts such as hooks with LPs, combining and integrating AMM algorithms, and benchmarking the effects of oracles on AMMs. Also, analysis on functionality of of non-AMM-based DEX and hybrid models that are often used for derivative market purposes are worthwhile topics for further researches.